\documentclass[%
 reprint,
superscriptaddress,
amsmath,amssymb,
aps,
prb,
floatfix,
]{revtex4-1}

\usepackage{graphicx}		   
\usepackage{dcolumn}		   
\usepackage{bm}
\usepackage{gensymb}
\usepackage{float}

\begin{document}

\title{Momentum-resolved Electron Energy Loss Spectroscopy for Mapping the Photonic Density of States}

\author{Prashant Shekhar}
\affiliation{Department of Electrical and Computer Engineering, University of Alberta, Canada, T6G 2R3}
\author{Marek Malac}
\affiliation{National Institute of Nanotechnology, Alberta, Canada, T6G 2R3}
\author{Vaibhav Gaind}
\affiliation{KLA Tencor, Milpitas, California, USA}
\author{Neda Dalili}
\affiliation{National Institute of Nanotechnology, Alberta, Canada, T6G 2R3}
\author{Al Meldrum}
\affiliation{Department of Physics, University of Alberta, Canada, T6G 2R3}
\author{Zubin Jacob}
\affiliation{Purdue Quantum Center, College of Electrical and Computer Engineering, Purdue University, West Lafayette,  USA, IN 47907}
\affiliation{Department of Electrical and Computer Engineering, University of Alberta, Canada, T6G 2R3}
\email{zjacob@purdue.edu}


\begin{abstract}
Strong nanoscale light-matter interaction is often accompanied by ultra-confined photonic modes and large momentum polaritons existing far beyond the light cone. A direct probe of such phenomena is difficult due to the momentum mismatch of these modes with free space light however, fast electron probes can reveal the fundamental quantum and spatially dispersive behavior of these excitations. Here, we use momentum-resolved electron energy loss spectroscopy ($q$-EELS) in a transmission electron microscope to explore the optical response of plasmonic thin films including momentum transfer up to wavevectors ($q$) significantly exceeding the light line wave vector. We show close agreement between experimental $q$-EELS maps, theoretical simulations of fast electrons passing through thin films and the momentum-resolved photonic density of states ($q$-PDOS) dispersion. Although a direct link between $q$-EELS and the $q$-PDOS exists for an infinite medium, here we show fundamental differences between $q$-EELS measurements and the $q$-PDOS that must be taken into consideration for realistic finite structures with no translational invariance along the direction of electron motion. Our work paves the way for using $q$-EELS as the preeminent tool for mapping the $q$-PDOS of exotic phenomena with large momenta (high-$q$) such as hyperbolic polaritons and spatially-dispersive plasmons.
\end{abstract}

\maketitle

\begin{figure*}
  \includegraphics{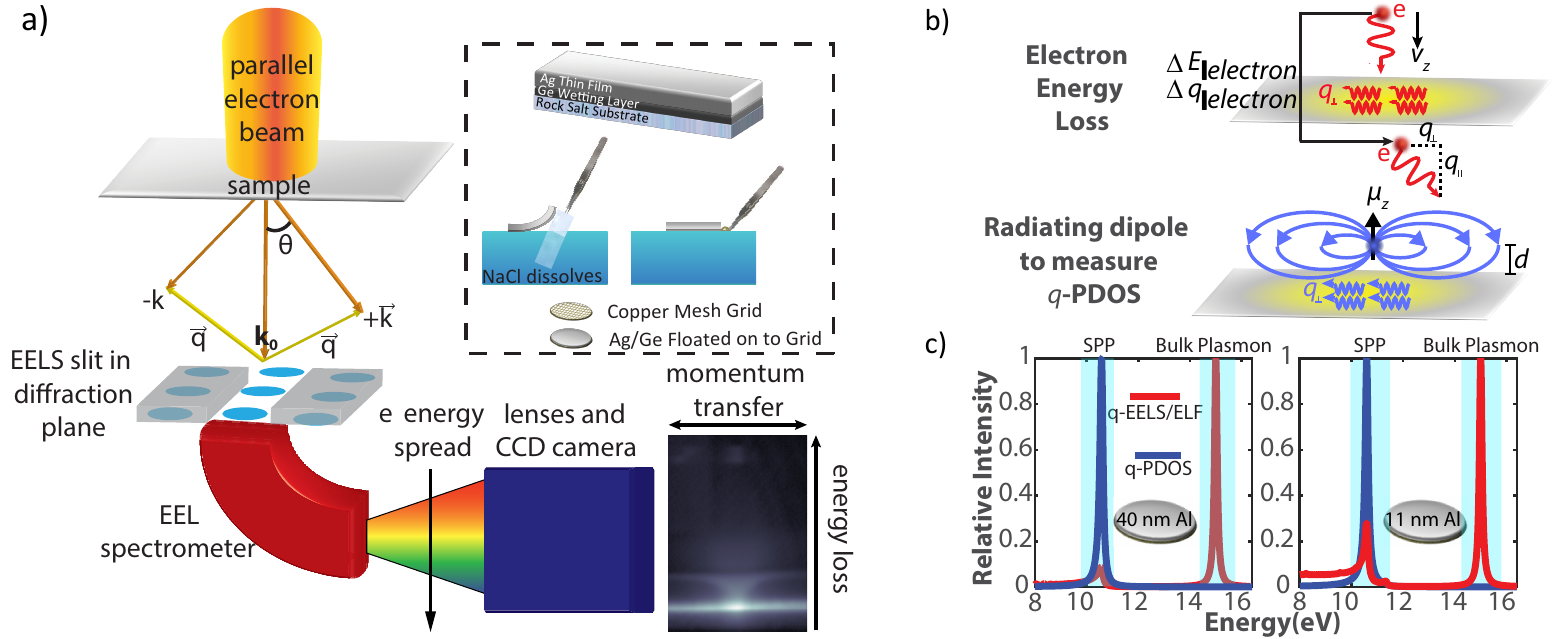}
  \caption{\textbf{$q$-EELS and $q$-PDOS} (a) The $q$-EELS experiment was performed with a Hitachi HF-3300 TEM with a GIF Tridiem\textsuperscript{TM} in $q$-EELS mode at 300 keV incident energy with parallel illumination resulting in a quantitative energy-momentum dispersion map of the excitations in the sample. The inset shows the sample preparation for an e-beam evaporated Ag thin film with a Ge wetting layer onto a copper mesh grid. (b) Schematic illustrating $q$-EELS with electron motion along the direction of no translational invariance (top) and a radiating dipole above a medium (bottom) for determining optical excitations in a material. For $q$-EELS, we consider normally incident electrons with velocity $v_z$ and probe momentum transfer parallel to the material interface ($\Delta q_\perp$) and energy loss ($\Delta E$) through the sample. The $q$-PDOS is measured by analyzing the power spectrum of a radiating dipole (with an oscillating source current) placed close to the material surface at a distance $d$. We only consider a dipole oriented perpendicular to the material interface (dipole moment $\mu$ only along $z$-direction).(c) The simulated relative $q$-EELS (determined by the energy loss function (ELF)) and the $q$-PDOS, integrated over the wavevector, for a 40 nm thick Al film (left) and a 11 nm thick Al film (right). The ELF is modeled for an electron with 300 keV incident energy while the $q$-PDOS is calculated for a radiating dipole 2 nm above the metal surface. For both thicknesses the ELF shows a strong peak at 15 eV corresponding to the bulk plasmon resonance of Al not seen in the $q$-PDOS. Both the $q$-PDOS and ELF show the surface plasmon polariton resonance at 10.6 eV. The aluminum is modeled with a simple Drude-like response with a plasma frequency ($\omega^{Al}_p$) of 15 eV and a damping factor ($\gamma^{Al}_p$) = 0.13 eV.}
  \label{fgr: IntDOS}
\end{figure*}

Electron energy loss spectroscopy (EELS) in a transmission electron microscope (TEM) is an essential tool for nanophotonics due to its ability to probe charge density oscillations far past the light-line. In EELS, a swift electron passes through a sample and experiences a measured energy loss ($\Delta E$) that corresponds directly to the transfer of the energy to characteristic excitations within the photonic nanostructure \cite{egerton_electron_2011}. Recently, scanning TEM EELS (STEM-EELS) has been used to spatially map plasmonic excitations on nanostructures with sub-nanometer spatial precision \cite{garcia_de_abajo_optical_2010,myroshnychenko_plasmon_2012,bosman_mapping_2007,nelayah_mapping_2007,colliex_electron_2016}, probe higher order modes of nanodisks \cite{hobbs_high-energy_2016} and nanoparticles \cite{raza_multipole_2015} as well as probe a series of phenomena interpreted to have quantum plasmonic behaviour \cite{ouyang_quantum_1992,scholl_observation_2013, zhou_atomically_2012,tan_quantum_2014}. However, in its current state, EELS does not provide a smoking gun for quantum excitations and similar experiments have also been described using the spatially dispersive properties of plasmonic excitations arising from the wavevector dependence of optical constants (non-local response). \cite{raza_blueshift_2013,wiener_electron-energy_2013,christensen_nonlocal_2014}. Additionally, EELS has been shown to provide insight into the nature of absorption versus scattering processes in nanostructures \cite{bernasconi_where_2017} as well as a direct relation to the photonic density of states (PDOS) \cite{garcia_de_abajo_probing_2008,ge_quasinormal_2016}.

Optical techniques, which use sources with small incident wavevectors, are severely limited in their ability to measure the PDOS at large wavevectors in photonic nanostructures \cite{garcia_de_abajo_optical_2010}. However, using electrons with techniques such as STEM-EELS and cathodoluminescence \cite{kuttge_local_2009}, this limitation can be surpassed as the inherently evanescent field of the electron can couple to large-wavevector excitations in the medium. Despite this, STEM-EELS provides no information about the band structure of the medium as the large spatial resolution achieved with the narrow beam fundamentally limits the momentum (angular) resolution possible with such a technique. This problem can be circumvented using momentum-resolved electron energy loss spectroscopy ($q$-EELS) where a wider parallel electron beam can measure both the transferred $\Delta E$ and momentum ($\Delta q$) from the electron to the sample to determine its characteristic energy-momentum dispersion relation \cite{batson_experimental_1983,pettit_measurement_1975} (Figure \ref{fgr: IntDOS}(a)). Thus, $q$-EELS is a valuable tool for the $q$-space mapping of the PDOS for plasmonic systems up to large wavevectors (high-$q$) and can give key insights into classical, quantum \cite{scholl_quantum_2012} and non-local optical phenomena from the measured band structure.

In this paper, we use $q$-EELS to measure the momentum-resolved photonic density of states ($q$-PDOS) of plasmonic excitations on ultra-thin silver films. We explore the role of electron energy and momentum loss as a function of thickness of the plasmonic film up to wavevectors 5 times past the light line. Although a direct connection between $q$-EELS and the $q$-PDOS has been theoretically proposed \cite{garcia_de_abajo_probing_2008},experiments confirming this phenomenon have been lacking. Also note that the relation between the two quantities have been determined for an optical source embedded in an infinite medium with translational invariance along the direction of electron motion. Thus, the established connection between $q$-EELS and the $q$-PDOS does not include the gamut of experimental systems with surface effects from finite structures integral to nanophotonics. Here, we highlight the fundamental differences between the $q$-PDOS and $q$-EELS in both energy and momentum space for such a finite system and experimentally demonstrate that $q$-EELS provides an accurate measure of the $q$-PDOS dispersion in energy-momentum space up to high-$q$ not possible with other techniques. We also conclude that coupling to longitudinal modes is not observed in the local model of the $q$-PDOS for an optical source placed outside the medium but is apparent in the $q$-EELS spectrum.  The use of $q$-EELS to map the $q$-PDOS to high-$q$ can pave the way for exploring more exotic phenomena such as hyperbolic polaritons \cite{shekhar_hyperbolic_2014,cortes_quantum_2012,poddubny_hyperbolic_2013}, slow light modes \cite{cortes_photonic_2013,baba_slow_2008} and strong coupling \cite{plumridge_ultra-strong_2008-1,shekhar_strong_2014}. It can also help shed light on questions related to non-local plasmonic excitations \cite{raza_nonlocal_2013} and the nature of nonclassical vs. classical effects \cite{yan_nonclassical_2016} effects in photonic nanostructures.

\section*{Results and discussion}
\label{scaling}

\textbf{Distinctions between the $q$-PDOS and $q$-EELS in Energy and Momentum Space.} The $q$-PDOS provides a framework that leads to a direct connection to Fermi's golden rule, making it a valuable tool for spontaneous and thermal emission engineering \cite{shekhar_hyperbolic_2014,cortes_quantum_2012,dyachenko_controlling_2016}. Here, we consider the $q$-PDOS for an optical source in vacuum above the medium of interest akin to many nanophotonic systems (Figure \ref{fgr: IntDOS} (b) bottom). It captures the near-field  interactions with photonic nanostructures from the power dissipated by a stationary oscillating electric dipole: $P = \frac{\omega}{2} Im(\mu^{*} \cdot \vec{E})$ where $\vec{E}$ is the electric field at the dipole position ($d$) produced by an oscillating current source $j_{pdos}(z,t) = -i\omega \mu e^{-i\omega t} \delta(z-d) \delta(x) \delta(y)$ \cite{ford_electromagnetic_1984} and $\mu$ is the dipole moment.

\begin{figure}
  \includegraphics{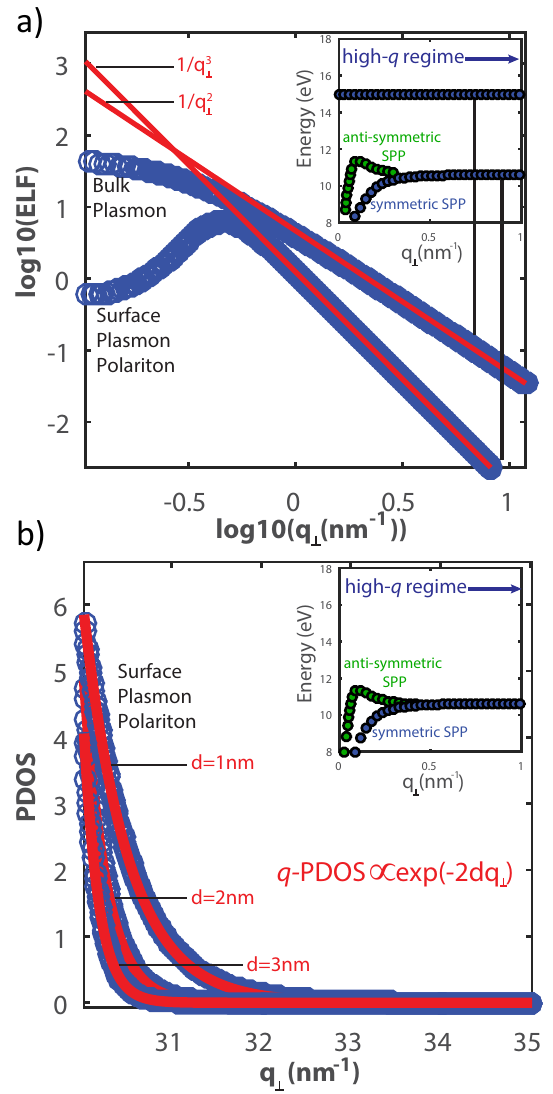}
  \caption{\textbf{$q$-PDOS and $q$-EELS Scaling with Wavevector} The scaling of the $q$-EELS (as determined by the ELF) (a) and $q$-PDOS (b) with respect to the wavevector parallel to the surface ($q_\perp$) is shown for an 11 nm thick Al film. At large wavevectors the ELF scales as $1/q^2$ and 1/$q^3$ for the bulk and surface plasmon polariton, respectively. The $q$-PDOS scaling with wavevector for the surface plasmon is seen to scale as $\exp(-2dq)$ where $d$ is the distance of the dipole from the top surface. The insets in (a) and (b) display the simulated $q$-EELS and $q$-PDOS dispersion, respectively. Note that both the $q$-EELS and $q$-PDOS show the symmetric and anti-symmetric surface plasmon in the band structure but only $q$-EELS shows the bulk plasmon dispersion at 15 eV.}
  \label{fgr: ScalingPlot}
\end{figure}

Although $q$-EELS measurements and the $q$-PDOS are comparable quantities, for a system with no translational invariance along the direction of electron motion, several key distinctions between the two quantities exist due to the different nature of their source excitations. In stark contrast to the stationary radiating dipole source above the medium in the $q$-PDOS, measurements made by $q$-EELS require a formalism for the scattering of a swift electron as it moves through matter. The energy loss and transferred momentum of an electron moving through a medium is described by the energy loss function (ELF) \cite{bolton_electron_1995} which is the work done by the retarding force of the fields induced ($E_{ind}$) by the electron: $U=\int d^3r \int dt E_{ind}(r,t)\cdot j_{eels}(r,t)$ where $r$ is the spatial position and $j_{eels}$ is the source current \cite{chase_electron_1970}. Note, unlike the oscillating current source in the PDOS ($j_{pdos}$), the source current in $q$-EELS is that of a moving charge: $j_{eels} = ev_z\delta(x)\delta(y)\delta(z-v_zt)$ where $v_z$ is the velocity of the electron perpendicular to the medium interface \cite{chase_electron_1970} (Figure \ref{fgr: IntDOS} (b) top). This contrasting nature of the source excitations for a finite structure consequently leads to fundamental variations between the $q$-PDOS and $q$-EELS (as determined by the ELF) in both energy and momentum space.

Figure \ref{fgr: IntDOS}(c) contrasts the $q$-PDOS and the ELF of an aluminum film as a function of film thickness and highlights a key difference between the two quantities in energy space: the local $q$-PDOS (integrated over the wavevector) for an emitter above the medium does not show any signature of the bulk plasmon resonance at 15 eV although it is a strong peak in the ELF for both the 11 nm and 40 nm thickness. Unlike a moving electron, the stationary radiating dipole source above the film has no longitudinal electric fields and therefore is unable to couple to any epsilon-near-zero resonances \cite{alu_epsilon-near-zero_2007,maas_experimental_2013} (bulk charge density excitations in a medium where the permittivity approaches 0) due to their longitudinal nature. Additionally, we observe that the ELF sees an increase in intensity at the surface plasmon polariton (SPP) energy (10.6 eV) relative to the bulk plasmon as the film thickness is decreased due to the electron probing  more effective surface compared to the bulk of the medium. This trade-off between the bulk and surface contribution to electron energy losses is known as the Begrenzungs effect \cite{egerton_electron_2011}. Although the ELF leads to a direct interpretation of the $q$-PDOS in energy space for an infinite medium, such intensity fluctuations of the surface plasmon as a function of film thickness do not occur in the local $q$-PDOS as it does not couple to the bulk plasmon for an emitter placed outside a finite structure.

We now turn our attention to the nature of the ELF and $q$-PDOS in momentum space with particular emphasis on the fundamentally different high-$q$ behaviour of plasmonic excitations. First, we consider the contribution to the $q$-PDOS ($\rho(\omega,d,q)$) for an emitter above a thin metal film from only the SPP (as there is no coupling to the bulk plasmon) and its dependence on the wavevector in the plane parallel to the material interface ($q_\perp$) \cite{ford_electromagnetic_1984}:

\begin{equation}
\rho_{spp}(\omega,d,q_{spp}) = \frac{\pi^2c^2}{\omega}Re\left (\frac{\sqrt{-\epsilon_m} q_{sp}^5}{1-\epsilon_{m}}e^{-2\sqrt{\frac{-1}{\epsilon_m}}q_{sp}d} \right)
\label{PDOS_Scale}
\end{equation}
where $\rho_{spp}$ is the surface plasmon contribution to the $q$-PDOS, $c$ is the speed of light in vacuum, $q_\perp=q_{spp}$ is the surface plasmon wavevector and $\epsilon_{m}$ is the permittivity of the metal. A similar expression can be derived for the ELF in the limit of high-$q$ for a thin metal slab surrounded by vacuum showing its dependence on the wavevector for both the bulk and surface plasmon contributions:

\begin{eqnarray}
ELF_{bulk} &=& \frac{t}{q_\perp^2}\left(\frac{v_z^2}{c^2}-\frac{1}{\epsilon_m} \right)\nonumber\\
ELF_{surf} &=& \frac{2}{q_\perp^3 \epsilon_m}\frac{\left[-f^2.(1+\epsilon_m)+\alpha-f(b^++b^-)\right]\alpha}{(1+\epsilon_m)^2f^2-\alpha^2}\nonumber\\
\label{ELF}
\end{eqnarray}
where $t$ is the slab thickness, $\alpha = (1-\epsilon_{m})$, $f=\exp({\sqrt{q_\perp^2-\epsilon_m\omega^2/c^2)}t})$, and $b^\pm=\exp(\frac{\pm\iota\omega t}{v_z})$. It is clear from equation \ref{PDOS_Scale} and equation \ref{ELF} that the scaling of the plasmonic excitations differ significantly for the ELF and $q$-PDOS intensity with respect to $q_\perp$. Figure \ref{fgr: ScalingPlot}(a) plots the ELF versus $q_\perp$ at the surface plasmon and bulk plasmon energy of Al in log scale. We note, that in the limit of large $q$, $ELF_{bulk}$ $\propto 1/q_\perp^2$ and $ELF_{surf}$ $\propto 1/q_\perp^3$. Conversely, at high-$q$, the PDOS scales such that PDOS $\propto\exp(-2dq_\perp)$ (Figure \ref{fgr: ScalingPlot}(b)). Thus, there is an increasing difference in momentum space between the ELF and the $q$-PDOS for finite structures as $q$ is increased that must be taken into consideration when performing $q$-EELS measurements.

\begin{figure*}
  \includegraphics{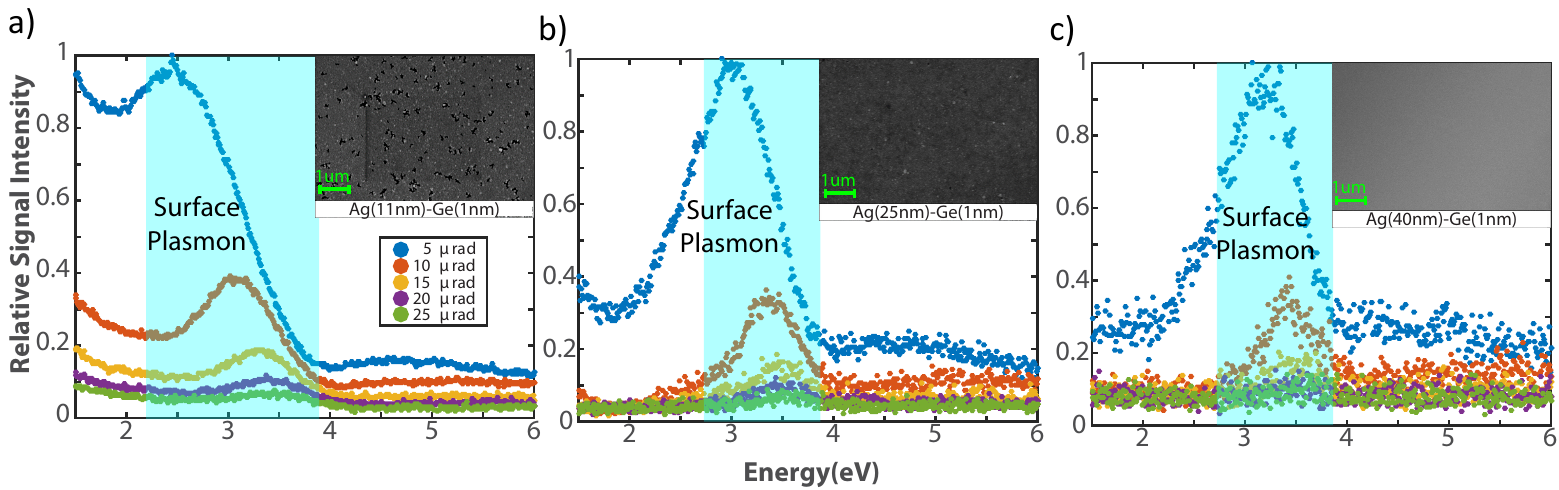}
  \caption{\textbf{$q$-EELS on Silver Films.} Relative experimental $q$-EELS scattering intensity at select scattering angles for an 11 nm (a) , 25 nm (b) and a 40 nm (c) Ag film. The film was deposited with 1 nm Ge wetting layer onto NaCl single crystals. A distinct peak (2.5 eV-3.5 eV) and a fainter peak at lower angles (4 eV-6 eV) correspond to the surface plasmon and the interband transitions of silver respectively. The inset is a scanning electron microscope image of the top surface of the silver film.}
  \label{fgr:Fig3}
\end{figure*}

\begin{figure*}
  \includegraphics{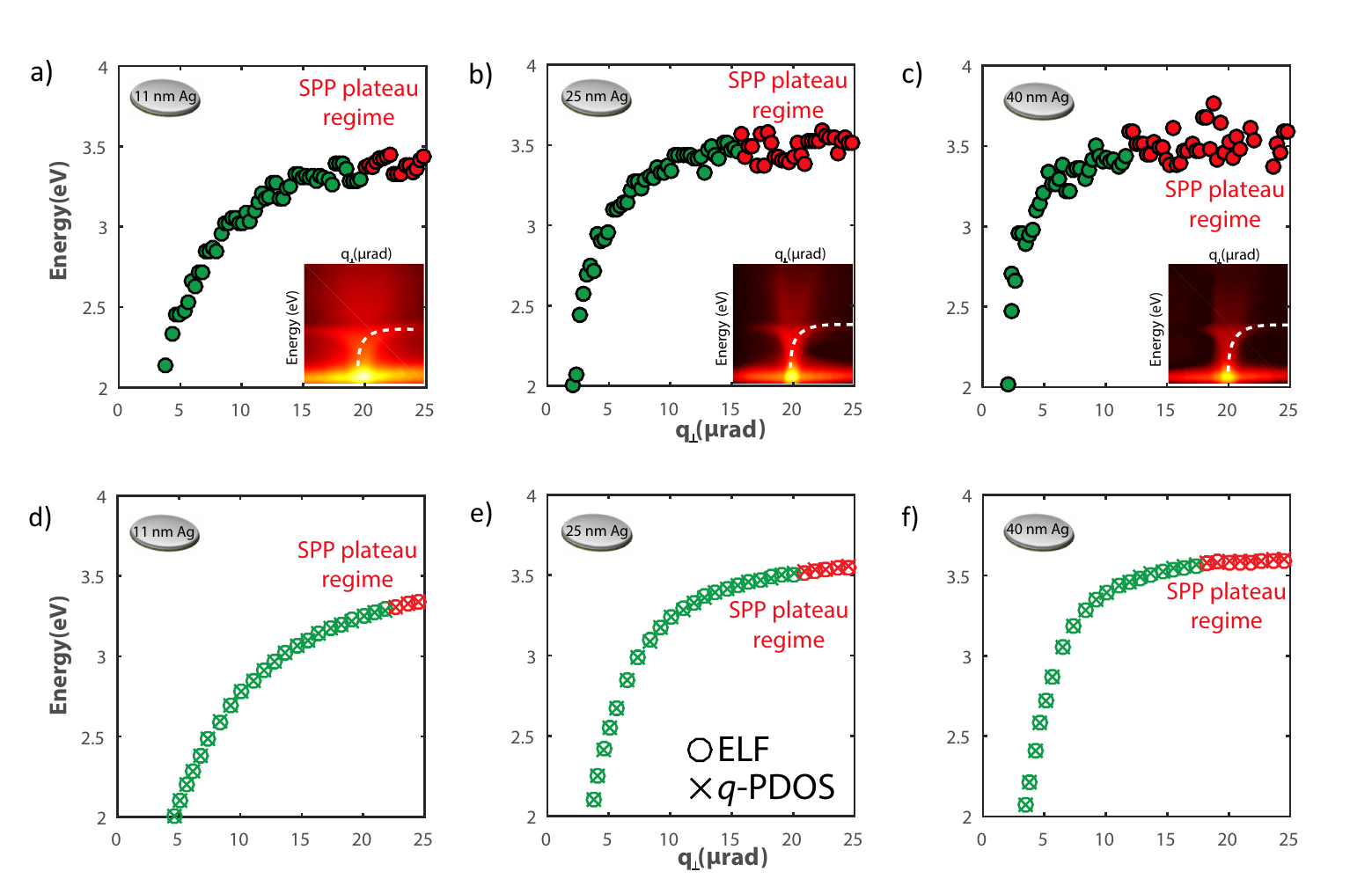}
  \caption{\textbf{$q$-PDOS Dispersion from $q$-EELS.} Experimental and theoretical $q$-EELS dispersion maps for an 11 nm (a,d), 25 nm (b,e) and a 40 nm thick (c,f) Ag film on a 1 nm Ge wetting layer. TOP: Energy-Momentum dispersion of the silver film from the raw experimental EELS data. A clear SPP dispersion is observed. Inset shows the generated experimental energy-momentum map with a dashed line indicating the SPP scattering intensity. Note the bright band at 0 eV in the inset corresponds to the zero-loss peak (ZLP). BOTTOM: Theoretical $q$-EELS scattering probability and $q$-PDOS for the various Ag films generating an energy-momentum map. A strong correspondence between the experimental and the simulated $q$-EELS and $q$-PDOS is observed for mapping the SPP dispersions. Note, that the SPP plateau appears at decreasing $q_\perp$ as the sample thickness increases in both theory and experiment.}
  \label{fgr:Fig4}
\end{figure*}

Although there exist some fundamental differences between $q$-EELS and the $q$-PDOS magnitudes for the system discussed above, once these theoretical differences are taken into account, $q$-EELS measurements can help to map the local $q$-PDOS as well as the energy-momentum band structure of plasmonic/polaritonic excitations. The insets of Figure \ref{fgr: ScalingPlot}(a,b) clearly highlight the ability of $q$-EELS to map the energy-momentum dispersion of the $q$-PDOS to great accuracy. The insets show the energy-momentum dispersion of the SPP, the anti-symmetric SPP, and, in the case of the ELF, the bulk plasmon for a 11 nm thick aluminum film. In the particular case of the SPP, both the $q$-PDOS and $q$-EELS show the gradual convergence of the SPP resonance to its plateau energy at 10.6 eV with one to one correspondence from the low-$q$ to high-$q$ regime.

\textbf{Dispersion mapping the $q$-PDOS with $q$-EELS.} In this section, we perform $q$-EELS as a function of film thickness to determine the $q$-PDOS dispersion of the SPP. We fabricated 11 nm, 25 nm and 40 nm continuous large grain sized free standing silver films. Note, while analysis with Al films was considered in the previous sections to highlight the effects of the bulk plasmon, we switch to Ag films in experiment for two key reasons: the wide use of Ag in nanophotonics systems due to plasmonic excitations in the visible regime and the fact that there are no bulk plasmon contributions for Ag close to the SPP energy. Detailed experimental methods, including fabrication of free standing Ag films and the $q$-EELS specifications, are outlined in the Methods section.

Figure \ref{fgr:Fig3} shows the experimentally measured relative $q$-EELS scattering probability at different scattering angles (corresponding to transferred momentum $q$) for an 11 nm, 25 nm and 40 nm thick Ag film on a 1 nm Ge wetting layer. The insets in the top row of Figure \ref{fgr:Fig4} (a, b, c) show the raw experimental $E$-$q$ dispersion map with energy loss in eV and momentum transfer in $\mu$rad. The intense band evident at 0 eV across all scattering angles is the zero-loss-peak (ZLP) representing unscattered and elastically scattered fast electrons present in all $q$-EELS spectra. The bright band at $\approx$3-3.5 eV (marked by the dashed white line) is the SPP peak of Ag and the series of bands in the 4-6 eV range evident at lower scattering angles ($\approx$5-10$\mu$rad) correspond to the interband transitions in Ag. Figure \ref{fgr:Fig3} is plotted by taking 1D line profiles along the designated scattering angles of the $E$-$q$ map. The strongest peak in the experimental energy loss spectra is that of the surface plasmon of silver as is expected for relatively thin films ($<$ 100 nm thick) where surface loss contributions dominate bulk losses. The relative scattering intensity of the surface plasmon also decreases with increasing transferred momentum for all thicknesses as expected due to the scaling of the ELF with $q$ observed in Figure \ref{fgr: ScalingPlot} (a) and equation \ref{ELF}.

Direct proof of the ability of $q$-EELS to map the $q$-PDOS dispersion from low-$q$ to high-$q$ is demonstrated in Figure \ref{fgr:Fig4} as seen by the strong match between the $q$-EELS experiment, ELF and the $q$-PDOS while mapping the SPP dispersion of Ag. (d), (e), and (f) show the near perfect agreement between the theoretical ELF and the $q$-PDOS across all thicknesses and $q$ implying the ability of the ELF (and therefore $q$-EELS measurements) to map the $q$-PDOS dispersion to high-$q$. This is further corroborated by the experimental $q$-EELS results shown in (a), (b) and (c) which shows a strong correspondence with the theory. Not only do the $q$-EELS measurements and ELF capture the broad $q$-PDOS dispersion, but also the nuanced changes in the SPP dispersion as the film thicknesses is increased. This is evident as the SPP dispersion profile for the 11 nm film (Figure \ref{fgr:Fig4} (a,d)) is shifted to higher momentum at lower energies and shows a more gradual convergence to the surface plasmon plateau energy (3.5 eV) than either the 25 nm (Figure \ref{fgr:Fig4} (b,e))  or 40 nm (Figure \ref{fgr:Fig4} (c,f)) film in both theory and experiment. The slight shift of the SPP dispersion to lower momentum by $\approx$2 $\mu$rad in experiment versus simulation is likely due to oxidation of the Ag film not included in the simulation.

In conclusion, despite being fundamentally different quantities for realistic finite structures with no translational invariance along the path of electron motion, $q$-EELS is a valuable tool for mapping the $q$-PDOS dispersion in photonic nanostructures from the low-$q$ to high-$q$ regime not possible with other techniques. The versatility of the $q$-EELS approach allows for mapping the $q$-PDOS dispersion for a wide variety of photonic nanostructures including photonic crystals, 2D materials, metamaterials, and metasurfaces including periodic arrays of structures composed of the wide array of nano plasmonic antennas. However, for periodic structures, the interplay between the periodicity, angular extent of the zero loss peak and the dynamic range of the $q$-EELS spectrum has to be optimized. Thus, $q$-EELS is a valuable tool for the $q$-space engineering of many exotic phenomena in nanophotonics including Cherenkov radiation \cite{yurtsever_formation_2008-1}, slow-light modes \cite{cortes_photonic_2013,baba_slow_2008}, non-local plasmonic excitations \cite{raza_nonlocal_2013}, hyperbolic modes \cite{shekhar_hyperbolic_2014,cortes_quantum_2012,poddubny_hyperbolic_2013}, and strong coupling \cite{plumridge_ultra-strong_2008-1,shekhar_strong_2014}.

\section*{Methods}
\label{Experiment}

Smooth, thin film samples with continuous and large grains are needed for $q$-EELS measurements. Such films limit the scattering of valence electrons from grain boundaries and the surface of the film, reducing the spurious background and improving momentum resolution. Additionally, the films must be deposited on soluble substrates, such as NaCl, in order to make the films free-standing to allow the fast electrons in the TEM to pass through the sample. Smooth 11 nm, 25 nm and 40 nm thick polycrystalline silver films were prepared by electron beam evaporation onto NaCl substrates with a 1 nm Ge wetting layer {\cite{chen_ultra-thin_2010} (FESEM images in Figure \ref{fgr:Fig3} insets). The NaCl substrates, with (100) orientation, were freshly cleaved less than 1 minute before they were placed in a vacuum chamber. High purity 99.999\% Ag and Ge sources were evaporated at ambient temperature $(12^{o}C-18^{o}C)$ under high vacuum ($8\times 10^{-7}$ torr) at  1\AA/s and 0.1\AA/s respectively. The samples were then floated off the substrate onto a TEM grid (inset Figure \ref{fgr: IntDOS}(a)) and inserted into the Hitachi HF-3300 TEM that has pressures $<5\times10^{-8}$ torr measured near the specimen. The sample was exposed to atmosphere for approximately 20 minutes during the float off process.

Performing $q$-EELS requires a notably different setup of the TEM compared to momentum-integrated EELS or STEM-EELS techniques (Figure \ref{fgr: IntDOS} (a)). Here, $q$-EELS was conducted with a Hitachi HF-3300 TEM/STEM with a cold field emission gun (CFEG) and a Gatan Image Filter (GIF) Tridiem\textsuperscript{TM} and the MAESTRO central computer control system \cite{bergen_centralized_2013}. The TEM operation in $q$-EELS uses a parallel electron beam (300 keV incident energy), unlike the point like probe of STEM-EELS with a highly convergent beam, in order to map $q$-space dispersion of the excitations. Electrons with normal incidence pass through the sample and are scattered with a momentum transfer ($\Delta q$) and undergo an energy loss ($\Delta E = \hbar\omega$) corresponding directly to the momentum and energy of excitations in the sample with resolutions of $\approx$ 0.35 $\mu$rad and $\approx$ 0.30 eV, respectively down to $\approx$ 1.2 eV until the ZLP onset. A desired range of scattering angles (corresponding to transferred momentum q) is selected with an EELS slit in the diffraction plane and the high electron energies are dispersed using the EEL spectrometer.

The $q$-EELS experiment was performed in diffraction mode with a 3 meter camera length and the sample was illuminated with a 0.1 $\mu$m  diameter probe. The GIF was aligned using a series of energy selecting slits ranging from 10 eV to 2 eV and tuned to have non-isochromaticity to 1st and 2nd order well below tolerance (0.05 eV and 0.43 eV, respectively). Although the total GIF alignment was performed (including tuning for image distortions, achromaticity, and magnification), no energy selecting slit was used during the $q$-EELS acquisition. The parallel illumination allows for the entire $q$-EELS energy-momentum map image for each sample to be recorded using a 1 second acquisition time integrated over 5 images in the GIF spectroscopy mode. As the Ag thin films have isotropic plasmonic properties in $q$-space the direction of critical points of the Brillouin zone were not considered however they should be addressed for a non-isotropic plasmonic response. In addition, energy per pixel and momentum per pixel calibrations of the CCD camera were corroborated with a 200 nm thick silicon sample with a known lattice spacing.

\section*{acknowledgement}

The authors thank Ray Egerton and Sean Molesky for detailed EELS discussions, Yoshifum Taniguchi for guidance with the microscope optics, Mike Bergen for support using the MAESTRO central computer control system, and Ward Newman and Farid Kahlor for useful guidance and discussions regarding the thin film fabrication process. We also thank our funding sources NINT/NRC, NSF, NSERC and AITF.

\bibliography{EELS_v9}

\end{document}